# Biophysicochemical interaction of a clinical pulmonary surfactant with nano-alumina

F. Mousseau[1], R. Le Borgne[2], E. Seyrek[1] and J.-F. Berret[1*]

[1] *Laboratoire Matière et Systèmes Complexes, UMR 7057 CNRS Université Denis Diderot Paris-VII, Bâtiment Condorcet, 10 rue Alice Domon et Léonie Duquet, 75205 Paris, France*
[2] *ImagoSeine Electron Microscopy Facility, Institut Jacques Monod, UMR 7592, CNRS - Université Paris Diderot Paris-VII, Paris, France.*

We report on the interaction of pulmonary surfactant composed of phospholipids and proteins with nanometric alumina ($Al_2O_3$) in the context of lung exposure and nanotoxicity. We study the bulk properties of phospholipid/nanoparticle dispersions and determine the nature of their interactions. The clinical surfactant Curosurf®, both native and extruded, and a protein-free surfactant are investigated. The phase behavior of mixed surfactant/particle dispersions was determined by optical and electron microscopy, light scattering and zeta potential measurements. It exhibits broad similarities with that of strongly interacting nanosystems such as polymers, proteins or particles, and supports the hypothesis of electrostatic complexation. At a critical stoichiometry, micron sized aggregates arising from the association between oppositely charged vesicles and nanoparticles are formed. Contrary to the models of lipoprotein corona or of particle wrapping, our work shows that vesicles maintain their structural integrity and trap the particles at their surfaces. The agglomeration of particles in surfactant phase is a phenomenon of importance since it could change the interactions of the particles with lung cells.

# I - Introduction

Pulmonary surfactant, the fluid lining the epithelium of the lungs is a complex surface-active fluid that contains phospholipids and lipids (85% and 5%, respectively) and 10% proteins (SP-A, SP-B, SP-C, SP-D and serum proteins).[1-2] The biophysical functions of pulmonary surfactant are to prevent the collapse of small alveoli during expiration and the over-expansion of large alveoli during inspiration. It also preserves bronchiolar patency during normal and forced respiration.[1,3-4] Furthermore, it has an important immunological role of protecting the lungs from injuries and infections caused by inhaled particles, including micro-organisms, particulate matter or engineered particles.[5-10] More specifically, particles of sizes less than 100 nm end up significantly deposited in the alveoli, and are susceptible to interact with the lung fluid.[11-12]

To evaluate the risks of exposure to inhaled nanomaterials, recent studies have been focusing on the interaction of particles with membranes, more specifically on model systems made of DMPC (1,2-dimyristoyl-*sn*-glycero-3-phosphocholine) or DOPC (1,2-dioleoyl-*sn*-glycero-3-phosphocholine) unilamellar vesicles.[13-20] The review of the different interaction potentials between particles and membranes revealed the importance of the interplay between particle/vesicle attraction and bilayer bending energy.[17] For diameters lower than a critical size (order of 10 nm for silica), the particles decorate the outer surface of the membrane, and induce aggregation.[17-18] For larger particle diameters, supported phospholipid bilayers form and coat the particles.[13,16] In the latter case, it is suggested that the membrane invaginates and eventually engulfs the particle in a process that resembles endocytosis.[14,21] In some reports, this engulfment is dubbed ingestion or transmigration, because it involves the passage of the particle from the outer to the inner part of the object. Ingestion of latex particle (2 µm) was first evidenced by Dietrich *et al*. using time-lapse microscopy and single object manipulation.[13] Transmigration of silica-based nanoparticles (110 nm) into micron-sized liposomes combined with membrane invagination was evidenced by Le Bihan *et al*. using cryo-electron tomography.[16] Recent simulations studies have shown that the wrapping of bio-membranes around particles are favored if the adhesive interactions are sufficiently strong to compensate bending.[22-25] Simulations were also performed on hydrophobic nanoparticles and predicted the formation of a lipoprotein corona or insertion into the membrane.[22,24] In contrast, very little is known on the interactions between particles and membranes forming multi-lamellar vesicles and/or comprising different types of phospholipids and proteins. In this context, factors such as vesicle dispersity, variability of the molecular constituents, *pH* or salt concentration of the solvent have to be taken into account, and may lead to specific behaviors.

In this work, we use a clinical pulmonary surfactant, Curosurf® (Chiesi Pharmaceuticals, Parma, Italy) to evaluate the interactions of particles with membranes relevant to biology. Curosurf® is a commercially available surfactant developed as a medication for exogenous treatment of respiratory distress syndrome for premature infants. Pulmonary surfactant substitutes are either synthetic or derived from animals, and their clinical effects have been thoroughly documented.[23] As compared to other preparations, Curosurf® contains the highest amount of phospholipids and proteins, with concentrations of 76 g $L^{-1}$ of phospholipids and 0.45 (resp. 0.59) g $L^{-1}$ of SP-B (resp. SP-C) membrane proteins and has been shown to be an effective surfactant in the treatment of respiratory deficiency. It was shown that Curosurf® bulk solutions are dispersions of spherical multilamellar vesicles (MLVs) mainly composed of a mixture of phospholipids and surface-active proteins.[6,26] Bulk samples were examined using freeze-fracture and (cryo)-transmission electron microscopy to determine the size and shape distribution of MLVs, and the organization of bilayers in the presence of biological or synthetic additives.[27-29]

Previous studies focused on interfacial and rheological properties of surface layers, as it is relevant to its biophysical function in the lungs.[19,30-35] In the alveoli however, the pulmonary surfactant forms a layer of a few hundreds of nanometers,[1,36] and after crossing the air-liquid interface the nanoparticles will also interact with the phospholipid sub-phase. This secondary process could interfere with phospholipid bulk dynamics and exchanges, and perturb the surfactant equilibrium. Some studies of nanoparticles interacting with pulmonary surfactant bulk phases were reported, but their number is limited. These studies involved the formation of complexes formed at the air/liquid interface, and further analysis on the interactions of these complexes with lung epithelial cells.[26,37-41] Issues related to cytotoxicity, lung inflammation and ability to cause oxidative stress were also addressed. However, these studies do not focus primarily on the main driving forces of interaction.

Here we report on the bulk properties of a synthetic and of purified pulmonary surfactant, and highlight the nature of interactions of nanoparticles with phospholipid membranes. In a first part, the physico-chemical properties of pulmonary surfactant and particles are thoroughly characterized. Emphasis is put on stability, surface charge and ageing behavior of dispersions. We also use extrusion to prepare multi-lamellar vesicles of well-characterized size and dispersity. As a model of nanoparticle-pulmonary surfactant system, alumina ($Al_2O_3$) was chosen as it is one of the most widely used material in nanotechnology-based products and its toxicological relevance, especially for lungs, is already well-recognized and studied.[42-46] The formation of complexes between $Al_2O_3$-nanoparticles with pulmonary surfactant substitutes is demonstrated here. The complexes are the result of electrostatic interactions between oppositely charged vesicles and particles.

## II - Materials and Methods

### II.1 – Materials

Curosurf® (Chiesi Pharmaceuticals, Parma, Italy) is natural surfactant extract obtained from porcine lungs, and containing polar lipids such as phosphatidylcholine (about 70% of total phospholipid content), phosphatidylglycerol (about 30% of total phospholipid content) and about 1% of hydrophobic proteins SP–B and SP–C. It is suspended in 0.9% sodium chloride solution and appears as a white to creamy suspension. According to the manufacturer, its pH is adjusted as required with sodium bicarbonate to a pH of 6.2 in average, the actual pH being indeed comprised between 5.5 and 6.5 according to the different batches received.[31] Curosurf® was kindly provided by Mostafa Mokhtari (Kremlin-Bicêtre Hospital, Val-de-Marne, France) and by Ignacio Garcia-Verdugo (INSERM, Paris, France). Dipalmitoylphosphatidylcholine (DPPC) was obtained from Sigma-Aldrich, while 2-Oleoyl-1-palmitoyl-sn-glycero-3-phospho-rac-(1-glycerol) (POPG) and L-α-Phosphatidyl-DL-glycerol sodium salt from egg yolk lecithin (PG, Sigma-Aldrich, MDL number: MFCD00213550) were given by Ignacio Garcia-Verdugo from Institut Pasteur, Paris. Aluminum oxide nanoparticles (Disperal®, SASOL) were kindly given by Florent Carn (Laboratoire Matière et Sytèmes Complexes, Paris). The powder was dissolved in nitric acid (0.4 wt. %) and sonicated for 30 minutes to give suspensions at 10 g $L^{-1}$. Methanol, nitric acid (70%) and poly(diallyldimethylammonium chloride) (PDADMAC, $M_W$ < 100 000 g $mol^{-1}$) were purchased from Sigma-Aldrich. Water was deionized with a Millipore Milli-Q Water system. All the products were used without further purification.

### II.2 – Protein-free surfactant

Phospholipids DPPC, PG and POPG were initially dissolved in methanol, at 10, 10 and 20 g $L^{-1}$ respectively. These compounds were mixed in proper amounts for a final weight concentration of 80%/10%/10% of DPPC/PG/POPG. The solvent was evaporated under low pressure at 60 °C for 30 minutes. The lipid film formed on the bottom of the flask was then rehydrated with the addition of Milli-Q water at 60 °C and agitated at atmospheric pressure for another 30 minutes. Milli-Q water was added again to finally obtain a solution at 1 g L$^{-1}$.

**II.3 – Extrusion**
Extrusion of Curosurf® and protein-free surfactant was performed using an Avanti Mini Extruder (Avanti Polar Lipids, Inc. Alabama, USA). Solutions were prepared at 1 g L$^{-1}$ and extruded 50 times through Whatman Nucleopor polycarbonate membranes of different pore sizes (50, 100, 200 and 800 nm in diameter). For the interaction studies with nano-alumina, the MLV dispersions were then diluted to 0.1 g L$^{-1}$ and the $pH$ was adjusted to $pH5$ using 0.1 M hydrochloric acid.

**II.4 – Dynamic Light Scattering (DLS)**
Hydrodynamic diameters were measured using a Zetasizer Nano ZS equipment (Malvern Instruments, Worcestershore, UK). A 4 mW He-Ne laser beam ($\lambda$ = 633 nm) is used to illuminate sample dispersion and the scattered intensity is collected at an angle of 173 degrees. The second-order autocorrelation function is analyzed using the CONTIN algorithm to determine the average diffusion coefficient $D_0$ of the scatterers. Hydrodynamic diameter is then calculated according to the Stokes-Einstein relation, $D_H = k_B T / 3\pi\eta D_0$, where $k_B$ is the Boltzmann constant, $T$ the temperature and $\eta$ the solvent viscosity. Measurements were performed in triplicate at 25° C or 37° C after an equilibration time of 120 s.

**II.5 – Zeta potential**
Laser Doppler velocimetry was used to carry out the electrokinetic measurements of electrophoretic mobility and zeta potential with the Zetasizer Nano ZS equipment (Malvern Instruments, Worcestershore, UK). Measurements were performed 3 times, at 25 °C, after 120 s of thermal equilibration.

**II.6 – Optical microscopy**
Phase-contrast and bright field images were acquired on an IX71 inverted microscope (Olympus) equipped with 20×, 40× and 60× objectives. 30 µl of dispersion were deposited on a glass plate and sealed into a Gene Frame® (Abgene/Advanced Biotech) dual adhesive system. An EXi Blue camera (QImaging) and Metaview software (Universal Imaging Inc.) were used as the acquisition system. Silica glass slides were coated using poly(diallyldimethylammonium chloride) to improve the surface adhesion of the MLVs.

**II.7 – Transmission electron microscopy**
TEM imaging was performed with a Tecnai 12 operating at 80 kV equipped with a 1K×1K Keen View camera. For negative staining drops of suspensions were deposited on formvar-carbon coated 400 mesh copper grids. Negative stains were made with 1% uranyl acetate in water.

**II.8 – Interaction**

The interactions between lung fluid models and nanoparticles were investigated using a mixing protocol developed by us and had already been tested on a series of strongly interacting colloidal systems.[47-49] It was checked that this procedure provides reproducible results, in particular that the mixing order or concentration does not modify interactions. Batches of pulmonary phospholipid phases and nanoparticles were prepared in the same conditions of $pH$ ($pH$ 5) and concentration ($c$ = 0.1 g L$^{-1}$) and then mixed at different ratios, noted by $X = V_{Surfactant}/V_{NP}$ where $V_{Surfactant}$ and $V_{NP}$ denote the volumes of the phospholipid and particle solutions respectively. Because the concentrations of the stock solutions are identical, the volumetric ratio $X$ is equivalent to the mass ratio between constituents. This procedure was preferred to titration experiments because it allowed exploring a broad range in mixing conditions ($X = 10^{-3} - 10^{3}$), while keeping the total concentration in the dilute regime and using a low amount of pulmonary fluid and particle.[48,50-51] The interactions between the phospholipid vesicles and Al$_2$O$_3$ nanoparticles occurred rapidly, *i.e.* within a few seconds after mixing.

# III - Results and Discussion

## III.1 – Characterization and processability of Curosurf®

### III.1.1 – Structure of native Curosurf

Curosurf® solutions were prepared from the native dispersions at 76 g L$^{-1}$ by dilution using 5 mM phosphate buffer ($pH$ 6.4). Solutions at c = 0.1 and 1 g L$^{-1}$ were studied using optical microscopy and light scattering for characterization at room and body temperatures. Microscopy observations reveal the presence of micron sized objects with spherical symmetry undergoing rapid Brownian motion. Fig. 1a-d shows a selection of some of the Brownian particles, with sizes 0.4 µm, 0.8 µm, 1.6 µm and 3.2 µm respectively. An analysis of 100 objects resulted in a size distribution of median diameter 1.0 µm and standard deviation 0.5 µm (Fig. 1e). Referring to earlier studies, these particles are identified as uni- or multilamellar vesicles.[26-28,32] In this study, large membrane layers or tubular myelin commonly observed in the lung fluid were not observed by microscopy.[52] Dynamic light scattering performed on the same sample shows a bimodal distribution with two major peaks, at 80 and 800 nm (Fig. 1f). The DLS peak at 800 nm is in agreement with that found by microscopy. Light scattering also shows a population of small vesicles that could not be detected by microscopy. Data on the surfactant microstructure obtained at $T$ = 25 °C or $T$ = 37 °C, *i.e.* at temperatures where the membranes are in the gel or in the fluid phase respectively, were similar (see differential scanning calorimetry data in Supplementary Information, S1).

### III.1.2 - Extrusion

As discussed previously, Curosurf® dispersion in its native state is made of highly disperse MLVs. Concerning interactions with particles, dispersity is a major issue, in particular for data analysis and interpretation. To overcome this difficulty, extrusion was applied using polycarbonate membranes with different pore diameters of 50, 100 and 200 nm. For extrusion, native Curosurf® was first diluted to 0.1 g L$^{-1}$ at $pH$ 6.4, as in the previous section.

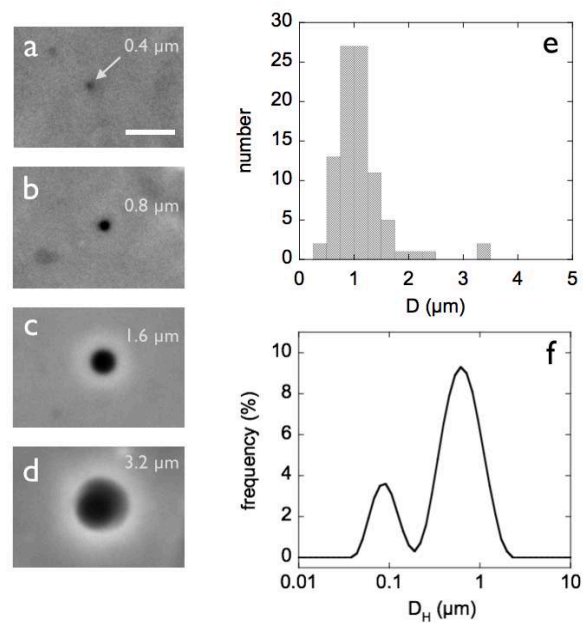

*Figure 1:* *a-d) Representative images of Curosurf® at the concentration of 0.1 g L$^{-1}$ obtained by phase contrast optical microscopy at T = 25° C (the bar is 3 μm). The sizes for the vesicles are indicated in the right hand side of the panel. e) Size distribution deduced from optical microscopy. f) Distribution of hydrodynamic diameters of Curosurf® vesicles as determined by dynamic light scattering (DLS).*

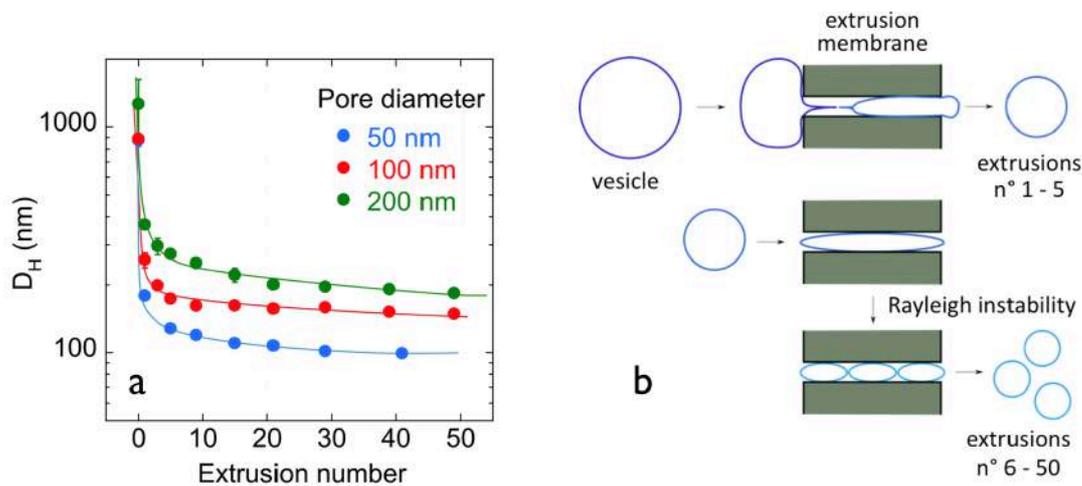

*Figure 2*: *a) Hydrodynamic diameter measured by light scattering as a function of the number of passages. Polycarbonate membranes of pore size 50, 100 and 200 nm were used and experiments were performed in triplicate. The error bars represent the standard deviation. For large extrusion numbers, the bars are smaller than the symbols. The solid lines are guides for the eyes. b) Schematic representation of the two-step extrusion process.*

Fig. 2a shows hydrodynamic diameter of vesicles as a function of the number of passages through the membrane. For the first extrusion passages (1 to 5), a sharp decrease in size was observed. With further extrusion, a plateau at hydrodynamic diameters 100, 150 and 180 nm is reached for membranes of increasing pore sizes. In parallel, the MLV size distribution was found to decrease progressively, reaching the final dispersity index of 0.1 ± 0.05 (S2). Additional extrusion using an 800 nm-pore membrane was performed for microscopic visualization purposes (see Section III.2).

Based on earlier studies of unilamellar vesicles,[53-54] a model was developed that can also describe the extrusion of pulmonary surfactant analogues. This model assumes that in the first passages, native micron size vesicles are pushed through the pores causing strong deformation, membrane tearing and resizing of the structure. At this stage, the exiting vesicles are still large and non-uniform. As extrusion progresses, the size mismatch between pores and vesicles lessens, giving rise to a Rayleigh-like instability. In the pores, the vesicles are sheared and stretched significantly, leading to their breaking into smaller objects (Fig. 2b). Because of this stretching, exit diameters are larger than the pore sizes. From purely geometrical considerations, the final diameters can be estimated and were found in agreement with experimental data (S2). Extrusion turns out to be an easy and reproducible procedure to prepare Curosurf® vesicles of predictable size.

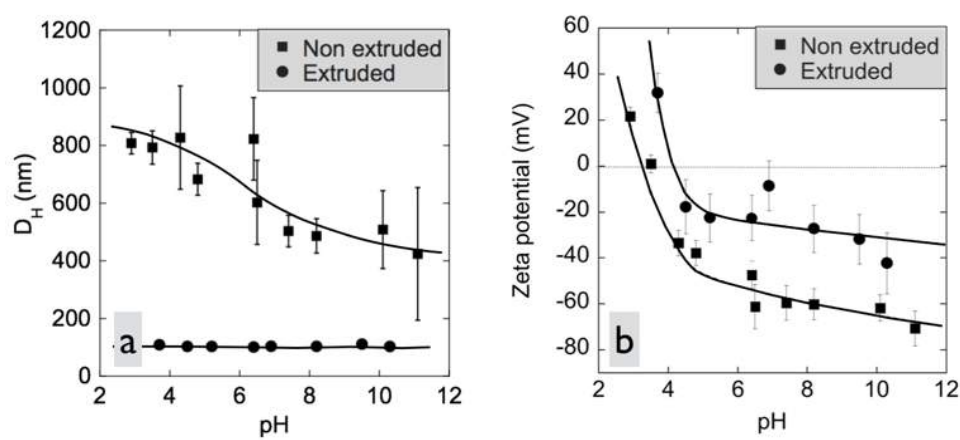

*Figure 3:* Hydrodynamic diameter (a) and zeta potential (b) as a function of $pH$ for extruded and non-extruded Curosurf® at 0.1 g $L^{-1}$. Error bars represent the standard deviation of experiments performed in triplicate. For extruded Curosurf® (close circles in Fig. 3a) the errors bars are smaller than the symbols. The solid lines are guides for the eyes.

### III.1.3 – Curosurf® stability: effect of $pH$ and ageing

From a practical standpoint and in anticipation of the work with particles, it is essential to know the behavior and stability of the surfactant vesicles as a function of physico-chemical parameters, such as concentration, temperature and pH. Hydrodynamic diameter and zeta potential of native and extruded Curosurf® were measured over a wide range of $pH$ above and below the physiological value ($pH$ 6.4 for newborns). Data are displayed in Figs. 3a and 3b respectively. With increasing $pH$, the broadly distributed vesicles of native Curosurf® decrease by 50% in size (from 800 to 400 nm). In contrast, the diameter of vesicles extruded through a 50 nm-membrane remains constant at 100 nm. Likewise, the zeta potential is found to decrease from positive values at acidic *pH* ($\zeta$ = + 25 mV) to highly negative ones at alkaline *pH* ($\zeta$ = - 30 mV and -60 mV for the extruded and native state respectively). At physiological conditions, we confirmed that phospholipid membranes are negatively charged.[55] The behaviors found in Fig. 3 can be understood in terms of acid-base properties of the phospholipid molecules. At *pH* below the pKa of the phospholipids (pKa of 4[56]), the phosphate moieties are not ionized and the overall charge of the membrane is given by the quaternary ammonium groups located in the zwitterionic heads. The rapid decrease of the $\zeta$-potential observed around *pH* 4 is then attributed to the deprotonation of the phosphate groups, and the charging of the membrane.

Ageing properties of Curosurf® preparations were investigated over a 30-day period by light scattering and zetametry. Experiments were performed every 12 hours, the samples being stored at 4° C between measures. Hydrodynamic diameter and zeta potential are displayed in Fig. 4a and 4b for native and extruded Curosurf® as well as for the protein-free surfactant for 7 days after the preparation. For size, native Curosurf® (either at 0.1 and 1 g L$^{-1}$) exhibits an evolution on the first day, but rapidly reaches a steady state. For these samples, the size distribution is broad and minor modifications could change the average hydrodynamic diameter. In contrast, the extruded samples remain unchanged over the entire period. The zeta potential shows a systematic increase from - 60 mV to - 30 mV for all samples, except for native Curosurf® at 1 g L$^{-1}$. Data at 7 and 30 days are moreover identical, confirming the good stability of as prepared MLVs. Concerning the origin of the $\zeta$-potential variations during the first days further work is required to clarify this issue. Additional results were acquired using up to four freeze-thaw cycles on the native specimens. No significant changes were observed either, suggesting that freezing surfactant preparation does not alter its structural properties (S3).

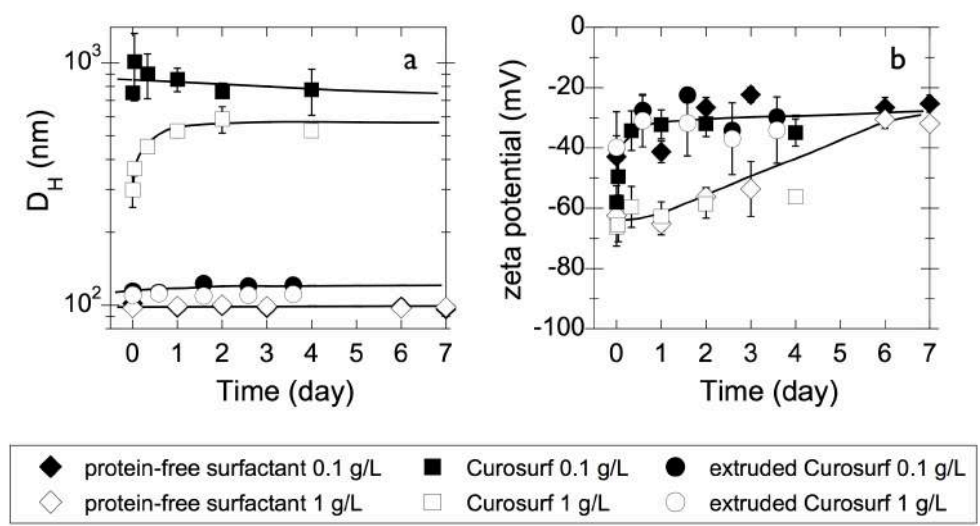

*Figure 4*: *Hydrodynamic diameter (a) and zeta potential (b) as a function of time for three types of lung fluid substitutes: protein-free surfactant (diamond), extruded (circles) and native (squares) Curosurf® at concentrations 0.1 and 1 g L$^{-1}$. Experiments were performed in triplicate and the error bars represent the standard deviation. Continuous lines are guides for the eyes.*

## III.2 – Interactions with aluminum oxide particles

Prior to experiments with pulmonary surfactant, the alumina particles were also thoroughly characterized. Dynamic light scattering, TEM and zeta potential measurements were performed at concentrations 0.1 to 1 g L$^{-1}$ and from $pH$ 3 to $pH$ 12 (Figs. 5a and 5b). In the inset, a TEM image shows that Al$_2$O$_3$ particles are anisotropic, their shape is that of irregular platelets of average dimensions 40 nm long and 10 nm thick. The particles exhibit moreover a broad size distribution (Fig. 5a). Electrophoretic measurements show that the particles are positively charged at low pH, with $\zeta$ = + 40 mV. As $pH$ increases above 6, particle surface charge decreases progressively and reaches the isoelectric point around $pH$ 10. Beyond, the particles are negatively charged. These results are in agreement with those of Cabane and coworkers who reported an isoelectric point of 9.3 for alumina particles.[57-58] Above $pH$ 6, particles also start to aggregate and their hydrodynamic diameter increases, suggesting that the stability observed at low $pH$ is caused by electrostatic repulsion. Experiments involving pulmonary surfactant and particles were performed at $pH$ 5, where both dispersions are stable.

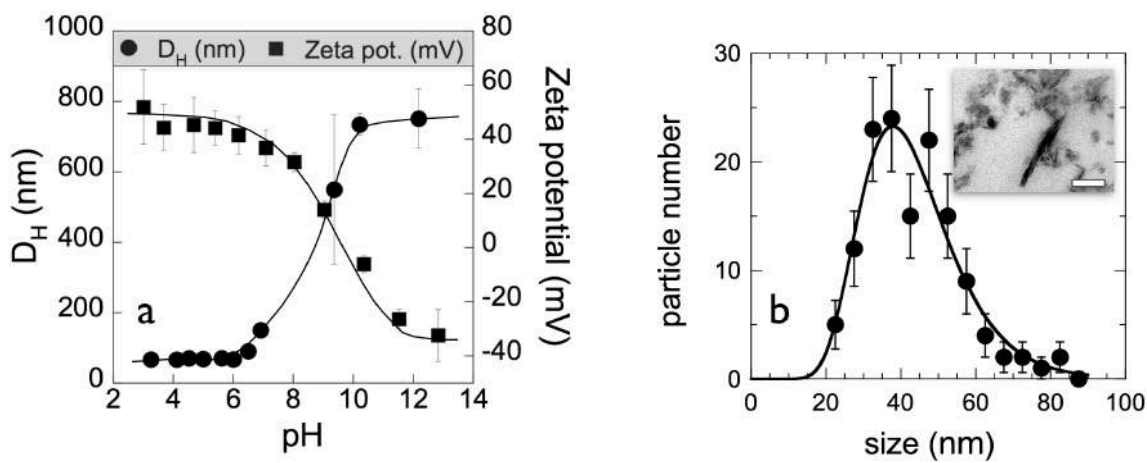

*Figure 5:* a) Hydrodynamic diameter and zeta potential of alumina particles (1 g L$^{-1}$) as a function of $pH$. Error bars represent the standard deviation of experiments performed in triplicate. b) Alumina particles size distribution obtained from transmission electron microscopy image analysis. Inset: TEM picture of Al$_2$O$_3$ particles (bar 50 nm). The solid line results from least-square calculations using a log-normal distribution with median diameter 41 nm and dispersity (defined as the ratio between the standard deviation and the average) of 0.3.

Surfactant/nanoparticle dispersions were obtained by mixing stock solutions at different volumetric ratios $X$. This method was successfully tested in the screening of multi-component phase diagrams.[48,51] In physiological conditions of lung exposure to nanoparticles, pulmonary surfactant is in excess and the ratio $X$ lies in the range $10^3 - 10^4$.[59] Because of this large excess, the dispersion is close to that of pure Curosurf®, and the interactions of the particles with the vesicles are difficult to assess. To remedy this shortcoming, the phase behavior of the mixed systems was investigated between $X = 10^{-3}$ and $10^3$. In this context, it is assumed that the nature of the interactions does not depend on the $X$-values. Fig. 6 shows the scattering intensity (6a to 6c) and the hydrodynamic diameter $D_H$ (6d to 6f) as a function of $X$ at $T = 25$ °C. The surfactant phases studied are a protein-free surfactant (Fig. 6a and 6d), and dispersions of extruded (using 50 nm pores in Fig. 6b and 6e) and native Curosurf® (Fig. 6c and 6f). The solutions were prepared at fixed concentration (0.1 g L$^{-1}$) for every $X$. In Figs. 6, nanoparticles and surfactant stock solutions are set at $X = 10^{-3}$ and $X = 10^3$ for convenience. For dilute solutions as the ones investigated here, the scattering intensity is proportional to the concentration and to the molecular weight of the scatterers. The continuous black lines in Figs. 6a, 6b and 6c are calculated assuming that surfactant and particles do not interact with each other, and that the scattering is the sum of the intensities weighted by their actual concentrations. In Figs. 6a-c, the scattering intensity is found to be systematically higher than the predictions for non-interacting species, implying that mixed aggregates are formed upon mixing. The scattering maxima ascertain moreover that the aggregates have a preferential stoichiometry, as the maximum is peaked at a definite $X$.

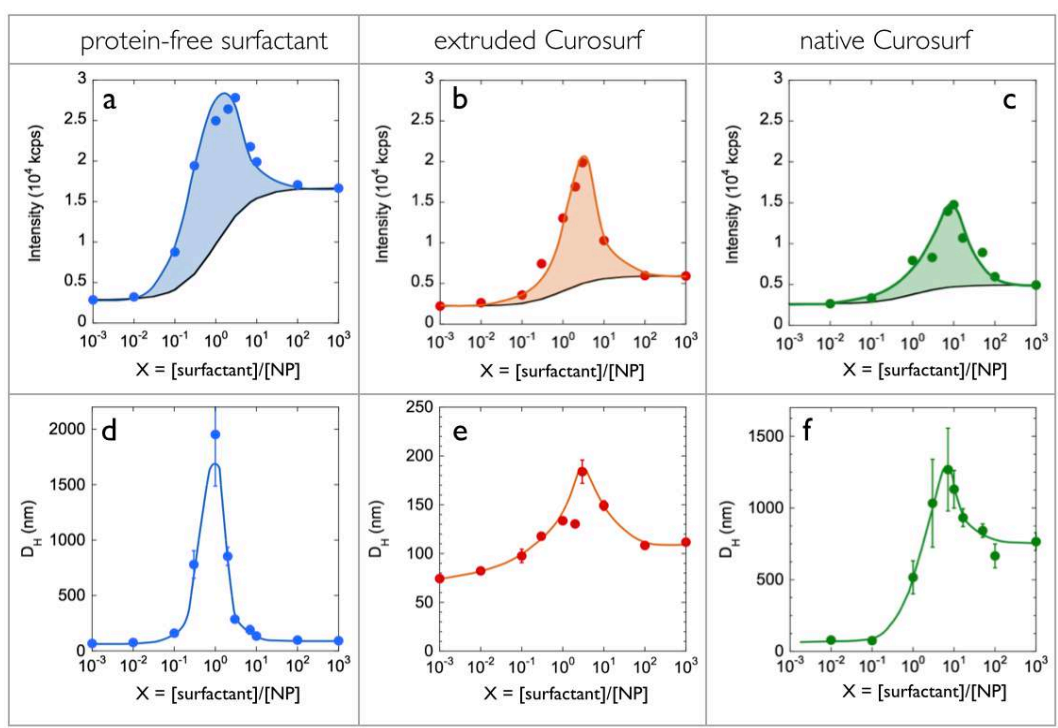

***Figure 6:*** *Scattered intensity (a, b and c) and hydrodynamic diameter (d, e and f) of alumina particles mixtures with protein-free surfactant (a, d), extruded Curosurf® (b, e) and native Curosurf® (c, f) as a function of X (T = 25 °C). X denotes the ratio between the surfactant and nanoparticles weight concentration. Light scattering experiments were performed in triplicate. The error bars represent the mean of the standard deviations. Solid lines in blue, red and green are guides for the eyes. Continuous lines in black in a, b and c represent the scattered intensity calculated assuming that particles and vesicles do not interact.*

The hydrodynamic diameters $D_H(X)$ shown in Figs. 6c-6e confirms this interpretation. The $D_H$-values at maximum reaches 2 µm for the protein-free surfactant, 200 nm and 1.5 µm for the extruded and native Curosurf®, and the position of the maxima corresponds to that of the scattering peaks. Similar results were obtained at T = 37 °C, *i.e.* at temperature where the phospholipids are in the disordered liquid state (S4). Fig. 7 displays the histograms of the interaction strength between particles and vesicles at the two temperatures, T = 25 °C and 37 °C. This strength is calculated from the integral of the scattering intensity (color-shaded areas in Fig. 6 and S4) with respect to the predictions for non-interacting species. This approach shows that surfactant-particle interactions are the strongest for protein-free surfactant, and the weakest for native Curosurf®. These conclusions were confirmed from the size data ($T = 25$ °C), which show increases by 1500%, 80% and 50% with respect to the uncomplexed vesicles. The shift of the scattering maxima for the different formulations (from $X = 1$ to 8 in Fig. 6a to 6c) is indicative of a predetermined stoichiometry between species. However, the number of vesicles and particles in the formed aggregate are difficult to estimate, since the relationship between the phospholipid concentration and the vesicle density is yet unknown.[60] This shift in $X$ suggests however that at a fixed concentration extruded vesicles are more numerous that native ones.

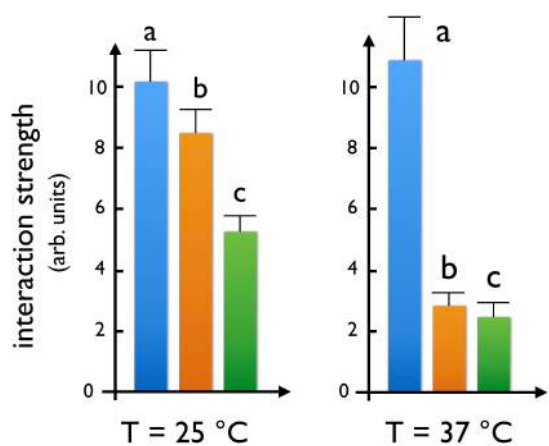

*Figure 7: Integrals over the mixing ratio X between the scattering intensities as determined experimentally and the intensities calculated from non-interacting particles and vesicles (colored areas in Fig. 6 and S4). The integral is indicative of the strength of nanoparticle-surfactant interactions. The labels **a**, **b** and **c** are related to the protein-free surfactant, the extruded Curosurf® and the native Curosurf®, respectively.*

To get further insight in the surfactant-$Al_2O_3$ mixed structures, dispersions were prepared at the maximum of the scattering peak using extruded Curosurf® with 100 nm or 800 nm pore membranes. Figs. 8a shows vesicles adsorbed at a silica surface before and after complexation with nano-alumina. In the inset, the uncomplexed 800 nm MLVs are isolated and randomly distributed on the substrate, whereas in the main frame they form micron-sized aggregates. These results are in agreement with the light scattering data (Fig. 6). Due to the sedimentation of the largest clusters, aggregates appear larger than those detected by light scattering. A close-up view of an aggregate shown in Fig. 8b exhibits characteristic patterns of vesicles at close packing inside the formed structure. Fig. 9a and 9b display aggregates observed by TEM, using here 100 nm extruded MLVs. The two examples shown are representative of the many structures observed using this technique. The cluster structure is dense and the vesicles have a doughnut shape. During the sample preparation, dehydration causes the collapse of the MLVs at the center. The vesicles maintain their shape on the border however due to the curvature effect of the phospholipid bilayer (Fig. 9c). Similar structures were observed on surfactant substitutes[28] and on polymer-liposome complexes.[29] Note that on the figure the $Al_2O_3$ particles are not visible, probably because of their weak contrast as compared to that of the stained vesicles. A schematic representation of an aggregate is illustrated in Fig. 9d.

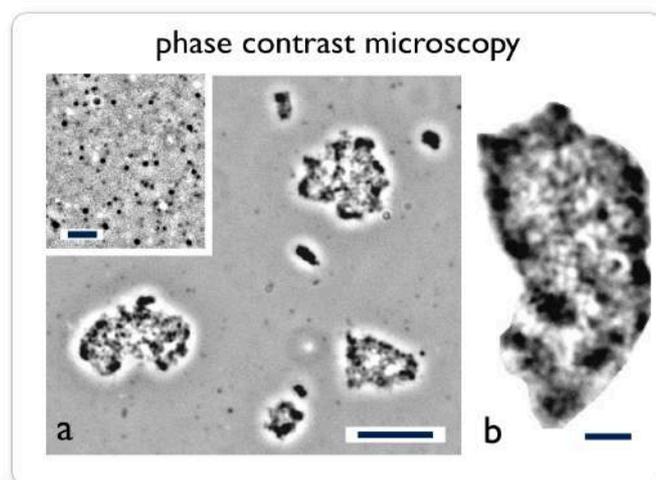

*Figure 8:* a) Dispersions of 800 nm Curosurf® vesicles and alumina particles at ratio $X = 7$ observed by phase contrast microscopy (×60). The bar is 20 µm. The sample was prepared at $c = 0.1$ g $L^{-1}$ and $T = 25$ °C and put between glass slides. The silica surfaces were coated using poly(diallyldimethylammonium chloride) to improve surface adhesion. Inset: image of uncomplexed vesicle in the same experimental conditions (bar 5 µm). b) Close-up view of an aggregate (bar 5 µm).

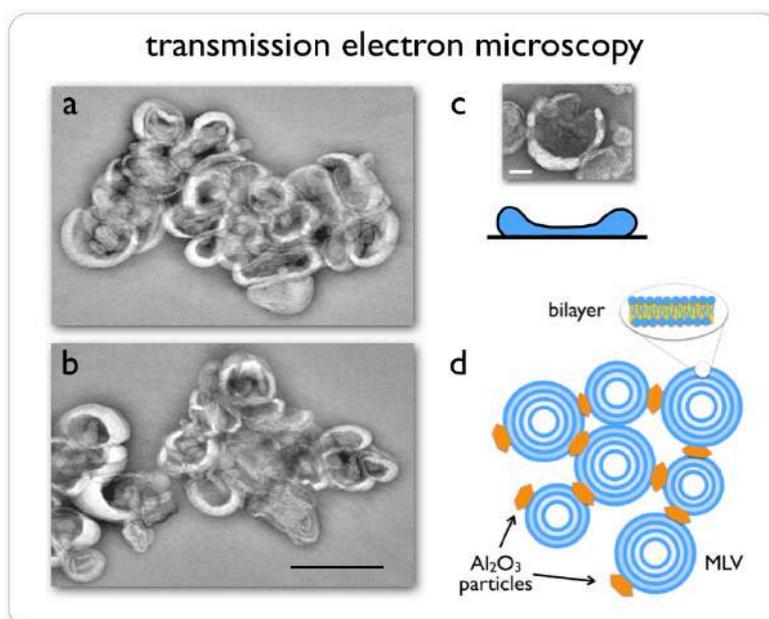

*Figure 9:* a and b) Transmission electron microscopy image of mixed 100 nm extruded Curosurf® vesicles and alumina particles (bar 200 nm). c) Close-up view of a single vesicle (bar 50 nm) and cartoon illustrating its doughnut height profile on the grid.[29] d) Schematic representation of nanoparticle-surfactant aggregates.

The results here demonstrate that the interaction of nano-alumina with surfactant gives rise to the formation of large aggregates, the size of which depends on the ratio between the two species. The phase behavior of the mixed dispersions exhibits similarities with that of strongly interacting nanosystems such as polymers, proteins or particles, and supports the hypothesis of electrostatic complexation. Electrostatic complexation designates the process by which co-assembly is driven by the pairing of electric charges located at the surfaces of particles or along the backbones of macromolecules.[49] It is concluded that, in the present experimental context the interactions of nano-alumina and surfactant are mainly driven by electrostatic forces. The destabilization or reorganization of the vesicles as a result of strong interaction is not observed. Encapsulation of alumina particles by vesicles is also not likely to occur. These scenarios would indeed result in a decrease in the light scattering data as a function of X, which is not observed experimentally. The interactions between nano-alumina and vesicles finally appear to be non-specific and independent of the membrane proteins SP-B and SP-C, as similar results were observed for protein-free surfactant and Curosurf®. At this point, it is difficult to attribute the differences between the protein-free surfactant and the Curosurf® (such as those seen in light scattering) to the presence of the proteins.

# IV - Conclusion

The pulmonary surfactant layer located at the air-liquid interface in the alveoli is known to be the first line of barrier for particles inhaled through the airways. Previous studies on the interaction with particles highlighted the modifications of interface properties. After deposition at the air-liquid interface however, the particles may dissolve in the surfactant sub-phase and modify the phospholipid exchanges and equilibrium. To investigate these interactions further, a clinical pulmonary surfactant Curosurf® and a synthetic analogue composed of phospholipids are characterized and analyzed for their interaction with 40 nm alumina nanoparticles. The size, the microstructure, the stability and the nature of the dispersion forces of the surfactant phase are the primary physico-chemical indicators to be considered when examining the interactions with nanoparticles. These properties were examined thoroughly in the present work. On native Curosurf®, data obtained by light scattering and microscopy corroborate those reported in the literature:[26,28,59] surfactant dispersions are made of multilamellar vesicles broadly distributed in size, typically from 100 nm to 5 μm. To circumvent the issue of the size dispersity, which can be detrimental in some cases (in particular in relation with data analysis and treatment), the Curosurf® dispersions were extruded using polycarbonate membranes of pores 50 to 800 nm, resulting in the generation of highly uniform vesicles. The first result that emerges from this work is that native and extruded Curosurf® behave similarly as a function of pH, temperature, stability and interactions with particles. We also demonstrate that for the bulk phase extruded surfactant represents an excellent model to study. With alumina particles, the interactions give rise to the formation of large aggregates made of the vesicles and particles, which is in agreement with the phenomenon of electrostatic complexation. The microstructure of the aggregates was disclosed using electron and optical microscopy, and it reveals densely packed clusters made of tens to hundreds of vesicles glued together by the nano-alumina (Fig. 9d). As anticipated, the behaviour of particles in Curosurf® bears some similarities with that of particles in biological fluids in general, such as plasma serum, lysosomal and interstitial fluids.[7,15,41,47,61] In these complex environments, the particles are generally found to be coated with proteins or with other biomacromolecules (forming then the protein corona), and later to agglomerate into clusters of various sizes.[62-63] In this respect, the vesicles of the lung surfactant substitutes are playing the same role as that of the serum proteins in other environments. The agglomeration of particles in biological fluids is a phenomenon of critical importance since it results in the loss of the nanometer character of the probes, in changes of hydrodynamic properties and interactions with cells. Finally, as compared to recent simulation predictions[22-25], our results provide a different view. The formation of large vesicular clusters is a process that is different from the phenomena of protein/lipoprotein corona or of formation of supported bilayers and should be considered when the biological responses are studied concerning the effects of nanoparticles in lungs. Moreover, aggregate formation could result in the long-term trapping of particles and prevent their interactions with epithelial cells.

## Acknowledgments

We would like to thank M. Angelova, A. Baeza-Squiban, F. Carn, I. Garcia-Verdugo, M. Mokhtari, F. Montel, D. Nguyen and N. Puff for fruitful discussions. We are also thankful to Lise Picaut from Laboratoire Matière et Systèmes Complexes for her help in the work on Curosurf® characterization and interaction, and Gaëlle Pembouong from Institut Parisien de Chimie Moléculaire for the DSC experiments. We acknowledge the ImagoSeine facility (Jacques Monod Institute, Paris, France), and the France BioImaging infrastructure supported by the French National Research Agency (ANR-10-INSB-04, « Investments for the future »). ANR (Agence Nationale de la Recherche) and CGI (Commissariat à l'Investissement d'Avenir) are gratefully acknowledged for their financial support of this work through Labex SEAM (Science and Engineering for Advanced Materials and devices) ANR 11 LABX 086, ANR 11 IDEX 05 02.

TOC Image for **"Biophysicochemical interaction of a clinical pulmonary surfactant with nano-alumina"**

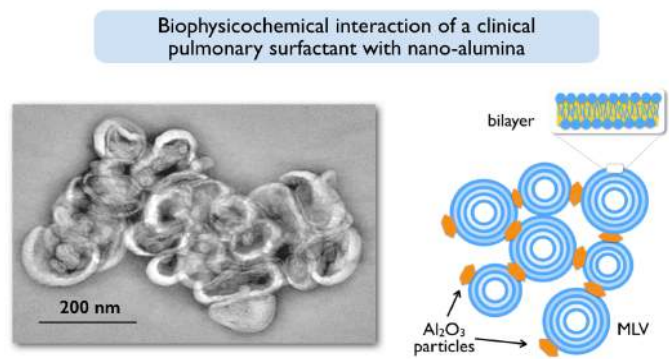